\documentclass[a4paper,11pt]{article}
\usepackage{pos}
\usepackage{subcaption}

\title{Investigating quark confinement from the viewpoint of lattice gauge-scalar models}
\author*[a]{Ryu Ikeda}
\author[a,b]{Kei-Ichi Kondo}

\affiliation[a]{Department of Physics, Graduate School of Science and Engineering, Chiba University,\\
  Chiba 263-8522, Japan}

\affiliation[b]{Department of Physics, Graduate School of Science, Chiba University,\\
  Chiba 263-8522, Japan}

\emailAdd{cdna0955@chiba-u.jp}
\emailAdd{kondok@faculty.chiba-u.jp}

\abstract{In this talk, first, we show that the color $N$-dependent area law falloffs of the double-winding Wilson loop averages for the $SU(N)$ lattice gauge model are reproduced from the $Z_N$ lattice Abelian gauge model due to the center group dominance in quark confinement.
Next, we discuss lattice gauge-scalar models which allow analytic continuation for gauge invariant operators between confinement region and Higgs region.
Applying the cluster expansion, we try to understand non-trivial contribution from scalar field in quark confinement mechanism.
In order to understand quark confinement further, moreover, we study double-winding Wilson loop averages in the analytical region on the phase diagram.}

\FullConference{%
 The 38th International Symposium on Lattice Field Theory, LATTICE2021
  26th-30th July, 2021
  Zoom/Gather@Massachusetts Institute of Technology
}

\begin{document}
\maketitle

\section{Introduction}

\begin{figure}[!h]
\centering
\begin{subfigure}{60mm}
  \centering\includegraphics[width=60mm]{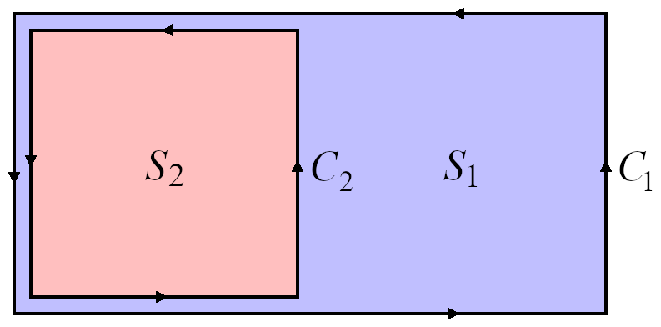}
  \hspace{0mm} \caption{  }
  \label{fig1}
\end{subfigure}
\begin{subfigure}{60mm}
  \centering\includegraphics[width=60mm]{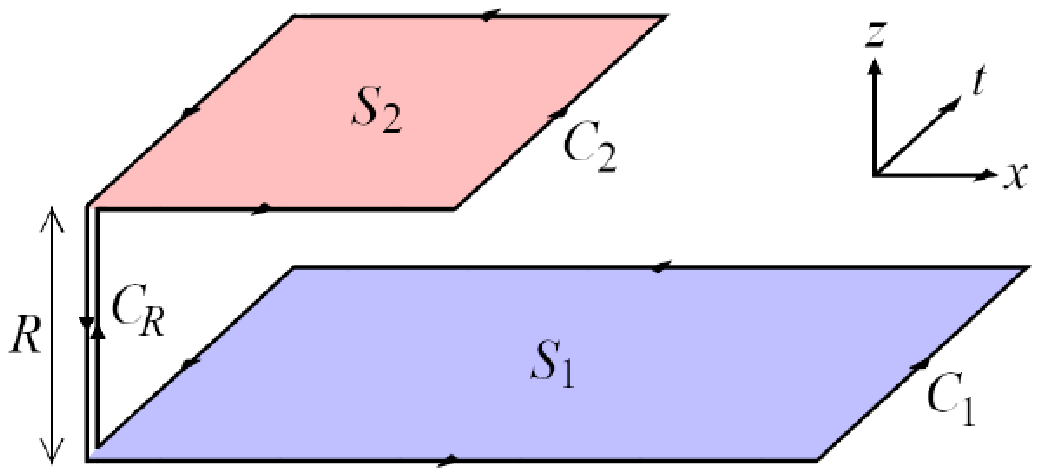}
  \hspace{0mm} \caption{  }
  \label{fig2}
\end{subfigure}
\hspace{0mm} \caption{(a) a ``coplanar'' double-winding Wilson loop, (b) a ``shifted'' double-winding Wilson loop.}
\label{fig12}
\end{figure}

In the lattice gauge theory, a double-winding Wilson loop operator $W(C_1 \cup C_2)$ has been introduced in \cite{GH15} to examine the possible mechanisms for quark confinement.
The double-winding Wilson loop operator is defined as a trace of the path-ordered product of gauge link variables ${U}_{\ell}$ along a closed loop $C$ composed of two loops $C_1$ and $C_2$:
\begin{align}
  W(C_1 \cup C_2) \equiv \mathrm{tr} \left[ \prod_{\ell \in C_1 \cup C_2} {U}_{\ell} \right] \ .
\end{align}
The double-winding Wilson loop is called \textit{coplanar} if the two loops $C_1$ and $C_2$ lie in the same plane, while it is called \textit{shifted} if the two loops $C_1$ and $C_2$ lie in planes parallel to the $x$-$t$ plane, but are displaced from one another in the transverse $z$-direction by distance $R$, and are connected by lines running parallel to the $z$-axis to keep the gauge invariance. See Fig.\ref{fig12}. Note that the double-winding Wilson loop operators are defined as a gauge invariant operator.

%%%%%%%%%%%%%%%%%%%%%%%%%%%%%%%%%%%%%%%%%%%%%%%%%%%%%%%%%%%%

%%%%%%%%%%%%%%%%%%%%%%%%%%%%%%%%%%%%%%%%%%%%%%%%%%%%%%%%%%%%

The area dependence of the expectation value $\langle W(C_1 \cup C_2) \rangle$ has been first investigated in \cite{GH15} to show that the coplanar double-winding Wilson loop average obeys the “difference-of-areas law” in the lattice $SU(2)$ Yang-Mills model by using the strong coupling expansion and the numerical simulations:
\begin{align}
  {\langle W(C_1 \cup C_2) \rangle}_{R=0} \simeq \exp [-\sigma ||S_1|-|S_2||] \ ,
\end{align}
where $S_1$ and $S_2$ are respectively the minimal areas bounded by loops $C_1$ and $C_2$. 

In the continuum $SU(N)$ Yang-Mills model, general multiple-winding Wilson loops have been investigated in \cite{MK17} to show that there is a novel “max-of-areas law” which is neither difference-of-areas law nor sum-of-areas law for multiple-winding Wilson loop average, provided that the string tension obeys the Casimir scaling for quarks in the higher representations.

In the lattice $SU(N)$ Yang-Mills model, it has been shown in \cite{KSK20} that the coplanar double-winding Wilson loop average has the $N$-dependent area law falloff in the strong coupling region:
“difference-of-areas law” for $N=2$, “max-of-areas law” for $N = 3$ and “sum-of-areas law” for $N \geqslant 4$:
\begin{align}
  {\langle W(C_1 \cup C_2) \rangle}_{R=0} \simeq
  \begin{cases}
    \exp [-\sigma ||S_1|-|S_2||] & (N=2) \\
    \exp [-\sigma \max (|S_1|,|S_2|)] & (N=3) \\
    \exp [-\sigma (|S_1|+|S_2|)] & (N \geqslant 4)
  \end{cases} \ .
\end{align}

Moreover, a shifted double-winding Wilson loop average as a function of the distance $R$ in a transverse direction has the long distance behavior which does not depend on $N$, while the short distance behavior depends on $N$.

In our investigation in \cite{IK21}, we examine the center group dominance for a double winding Wilson loop average. It has been shown in \cite{Frohlich79} that the ordinary single-winding Wilson loop average in the non-Abelian lattice gauge theory with the gauge group $G$ is bounded from above by the same Wilson loop average in the Abelian lattice gauge theory with the center gauge group $Z(G)$:
\begin{align}
 |\langle W_{R(G)}(C) \rangle_{G}(\beta)| \le 2 {\rm tr}({\bf 1}) \langle W_{R(Z(G))}(C) \rangle_{Z(G)}(2{\rm dim}(G)\beta) \ .
\end{align}

We have extended the above statement to the double winding Wilson loop average, beyond the case of the ordinary single-winding Wilson loop average:
\begin{align}
 |\langle W_{R(G)}(C_1 \cup C_2) \rangle_{G}(\beta)| \le 2 {\rm tr}({\bf 1}) \langle W_{R(Z(G))}(C_1 \cup C_2) \rangle_{Z(G)}(2{\rm dim}(G)\beta) \ .
\end{align}

From this point of view, we introduce the \textit{character expansion} to the weight ${e}^{S_G [U]}$ coming from the action and perform the group integration, in order to estimate the expectation value in the $Z_N$ lattice gauge model. We evaluate the double-winding Wilson loop average up to the leading contribution to show that the $N$-dependent area law falloff in the $SU(N)$ lattice gauge model can be reproduced by using the (Abelian) $Z_N$ lattice gauge model.
By taking the limit $N \to \infty$, we show the center group dominance for a double-winding Wilson loop average in the $U(N)$ lattice gauge model through the $U(1)$ lattice gauge model.

Finally, we extend the above arguments for the lattice gauge-scalar model on the “analytic region”. For this purpose, we estimate the area law falloff, the string tension, and the mass gap by using the \textit{cluster expansion}.

\section{Lattice $Z_N$ gauge model}
First, we consider the lattice $Z_N$ gauge model with the coupling constant defined by $\beta := 1/ g^2$ on a $D$-dimensional lattice $\Lambda$ with unit lattice spacing, which is specified by the action
\begin{align}
  S_G [U] = \beta \sum_{p \in \Lambda} \mathrm{Re} \ U_p \ , \quad U_p := \prod_{\ell \in \partial p} {U}_{\ell} \ ,
\end{align}
where $\ell$ labels a link, $p$ labels an elementary plaquette. To examine this $Z_N$ gauge model analytically, we introduce the \textit{character expansion} to the weight ${e}^{S_G [U]}$ to obtain the expanded form of the expectation value of an operator $\mathscr{F}$:
\begin{align}
  {\langle \mathscr{F} \rangle}_{\Lambda}
    &:= {Z}_{\Lambda}^{-1} \int \prod_{\ell \in \Lambda} d {U}_{\ell} \ {e}^{ S_G [U] } \mathscr{F}
    = {Z}_{\Lambda}^{-1} \int \prod_{\ell \in \Lambda} d {U}_{\ell} \ \prod_{p \in \Lambda} \sum_{n=0}^{N-1} b_n (\beta) U_p^n \mathscr{F} \ , \\
  {Z}_{\Lambda}
    &:= \int \prod_{\ell \in \Lambda} d {U}_{\ell} \ {e}^{ S_G [U] } \ ,
\end{align}
where the coefficients $b_n (\beta)$ is defined by
\begin{align}
  b_n (\beta) := \frac{1}{N} \sum_{\zeta \in Z_N} {\zeta}^{-n} {e}^{\beta \mathrm{Re} \ \zeta} \ .
\end{align}

We define $c_n (\beta) := b_n (\beta) / b_0 (\beta)$. For $N=2,3,4$ and $\infty$, $c_1 (\beta)$ and $c_2 (\beta)$ are written in the form
\begin{align}
  c_1 (\beta) &= \frac{ {e}^{\beta} - {e}^{-\beta} }{ {e}^{\beta} + {e}^{-\beta} } \quad (N=2) \ , \qquad 
  c_1 (\beta) = \frac{ {e}^{\beta} - {e}^{-\beta/2} }{ {e}^{\beta} + 2{e}^{-\beta/2} } = c_2 (\beta) \quad (N=3) \ , \notag \\
  c_1 (\beta) &= \frac{ {e}^{\beta} - {e}^{-\beta} }{ {e}^{\beta} +2+ {e}^{-\beta} } , \quad
  c_2 (\beta) = \frac{ {e}^{\beta} -2+ {e}^{-\beta} }{ {e}^{\beta} +2+ {e}^{-\beta} } \quad (N=4) \ , \notag \\
  c_1 (\beta) &= \frac{ I_1 (\beta) }{ I_0 (\beta) } , \quad
  c_2 (\beta) = \frac{ I_2 (\beta) }{ I_0 (\beta) } \quad (N=\infty) \ .
\end{align}

\begin{figure}[!h]
\centering
\begin{subfigure}{60mm}
  \centering\includegraphics[width=60mm]{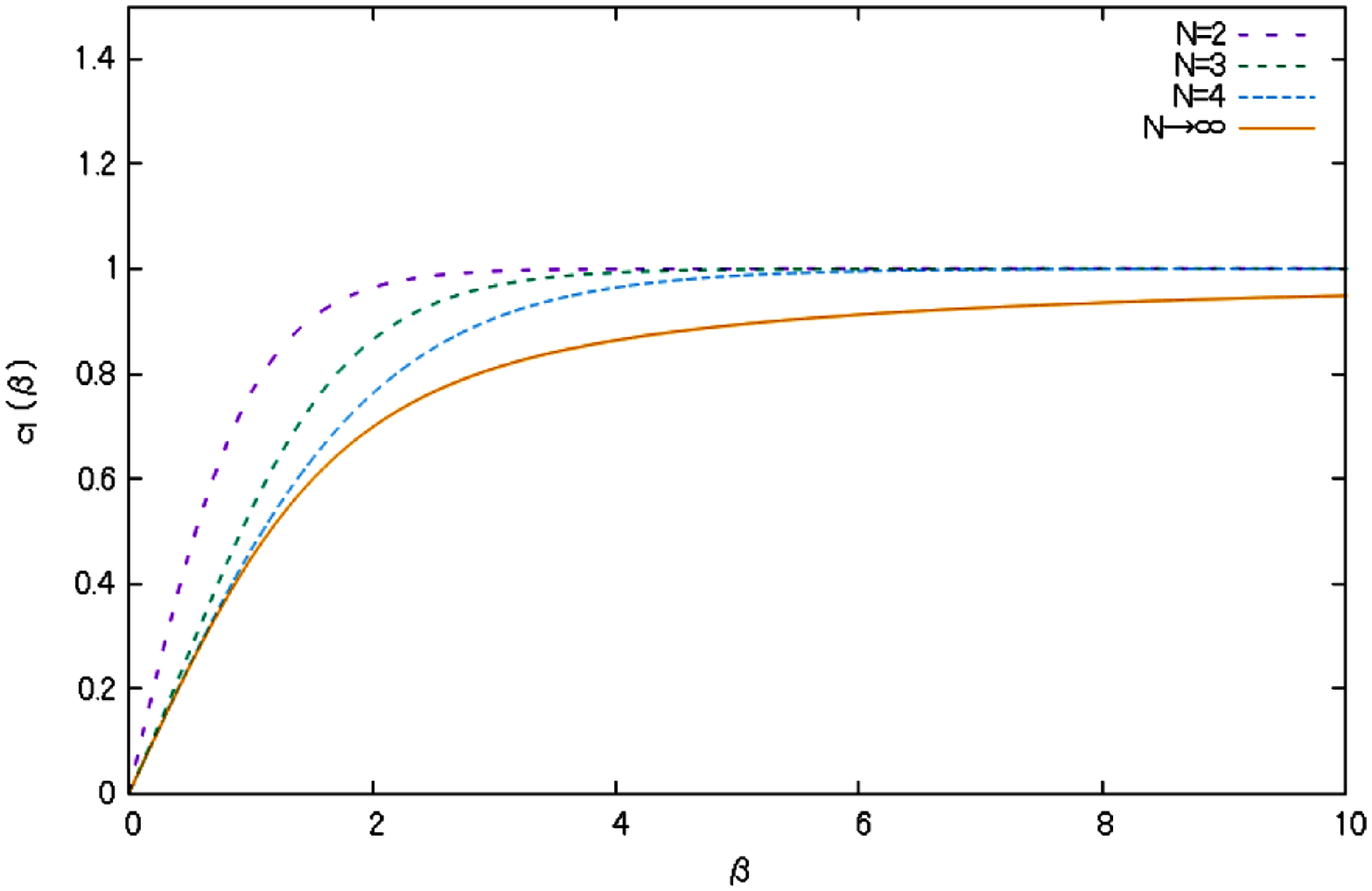}
  \hspace{0mm} \caption{$c_1 (\beta) \ (N=2,3,4,\infty)$}
  \label{fig3}
\end{subfigure}
\begin{subfigure}{60mm}
  \centering\includegraphics[width=60mm]{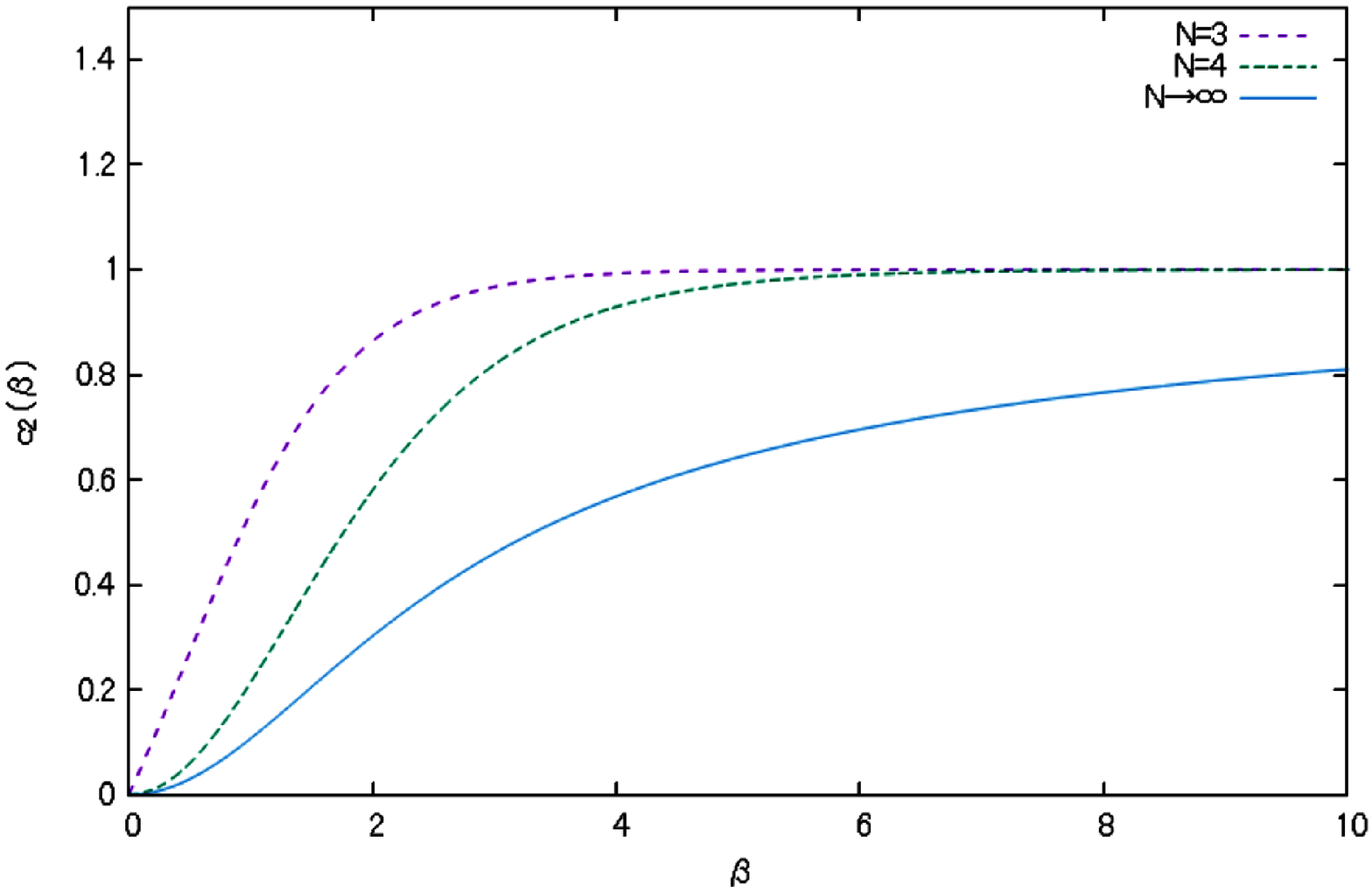}
  \hspace{0mm} \caption{$c_2 (\beta) \ (N=3,4,\infty)$}
  \label{fig4}
\end{subfigure}
\hspace{0mm} \caption{The character expansion coefficient as a function of $\beta$, (a) $c_1 (\beta)$, \ (b) $c_2 (\beta)$}
\label{fig34}
\end{figure}

Note that ${b}_{N-n} (\beta) = b_n (\beta)$ and $0 \leqslant c_n (\beta) < 1$ for $0 \leqslant \beta < \infty$. For $N=2,3,4$ and $\infty$ , the behavior of $c_1 (\beta)$ and $c_2 (\beta)$ as functions of $\beta$ are indicated in Fig.\ref{fig34}. We find that $c_1 (\beta) \sim \mathcal{O} (\beta) \ (N \geqslant 2)$ and $c_2 (\beta) \sim \mathcal{O} ({\beta}^{2}) \ (N \geqslant 4)$ for $\beta \ll 1$.

%%%%%%%%%%%%%%%%%%%%%%%%%%%%%%%%%%%%%%%%%%%%%%%%%%%%%%%%%%%%

%%%%%%%%%%%%%%%%%%%%%%%%%%%%%%%%%%%%%%%%%%%%%%%%%%%%%%%%%%%%

\begin{figure}[!h]
\centering
\begin{subfigure}{60mm}
  \centering\includegraphics[width=60mm]{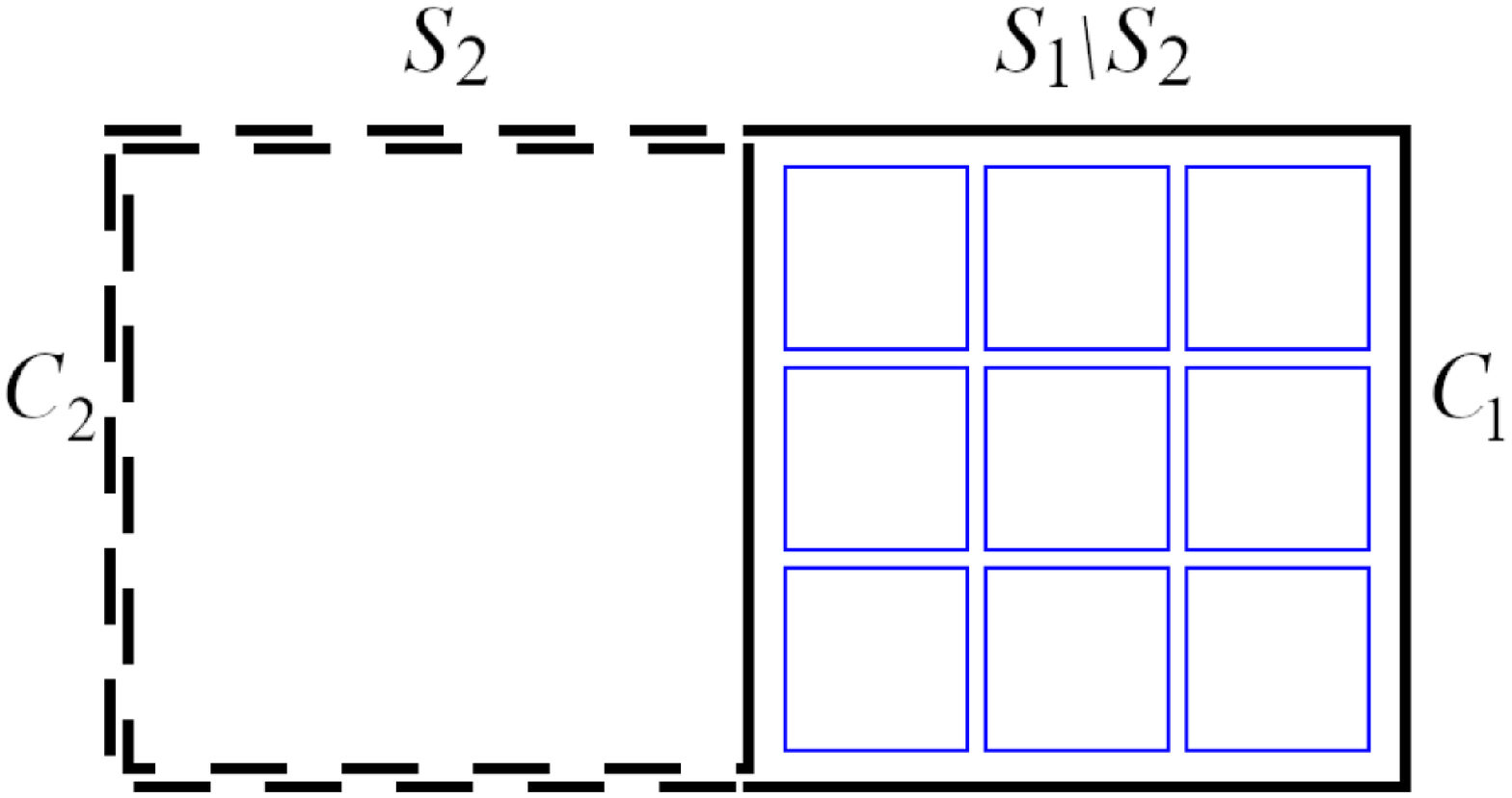}
  \hspace{0mm} \caption{$N=2$}
  \label{fig5}
\end{subfigure}
\begin{subfigure}{60mm}
  \centering\includegraphics[width=60mm]{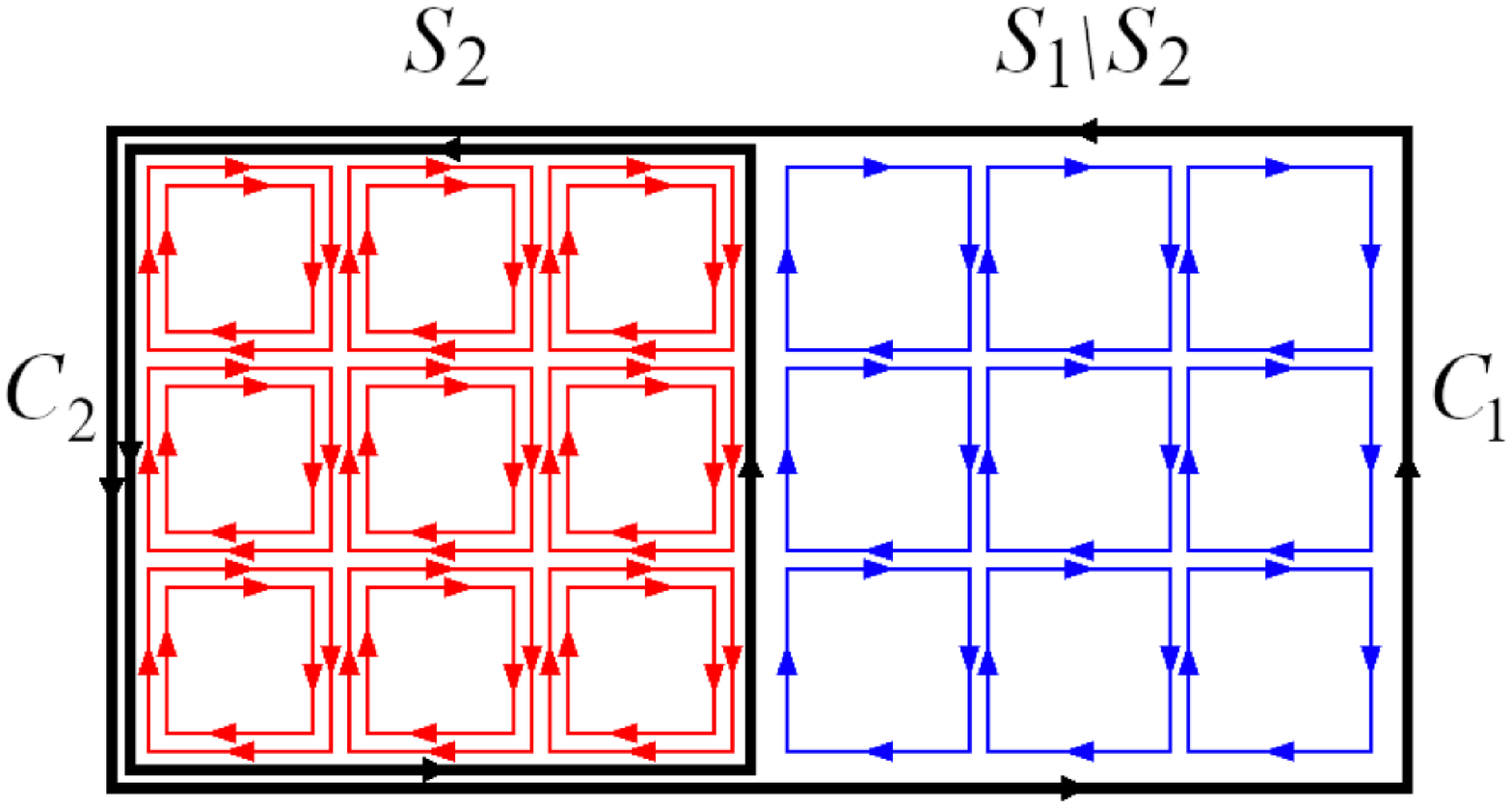}
  \hspace{0mm} \caption{$N \geqslant 3$}
  \label{fig6}
\end{subfigure}
\hspace{0mm} \caption{A coplanar double-winding Wilson loop, (a) $N=2$ , \ (b) $N \geqslant 3$}
\label{fig56}
\end{figure}

Next, we evaluate the expectation value of a coplanar double-winding Wilson loop in the lattice $Z_N$ pure gauge model.
The leading contribution to a coplanar double-winding Wilson loop average is given by the tiling of a planar set of plaquettes, as shown in the Fig.\ref{fig56}. (These result are exact for all $\beta$ when $D=2$, while valid for $\beta \ll 1$ when $D>2$.)

%%%%%%%%%%%%%%%%%%%%%%%%%%%%%%%%%%%%%%%%%%%%%%%%%%%%%%%%%%%%

%%%%%%%%%%%%%%%%%%%%%%%%%%%%%%%%%%%%%%%%%%%%%%%%%%%%%%%%%%%%

The result of the coplanar double-winding Wilson loop average up to the leading contribution is given by
\begin{align}
  {\langle W(C_1 \cup C_2) \rangle}_{R=0} \simeq
  \begin{cases}
    {c_1 (\beta)}^{|S_1|-|S_2|} & (N=2) \\
    {c_1 (\beta)}^{|S_1|} & (N=3) \\
    {c_2 (\beta)}^{|S_2|} {c_1 (\beta)}^{|S_1|-|S_2|} & (N \geqslant 4)
  \end{cases} \ .
\end{align}
Then we obtain the (non-zero) string tension from this result:
\begin{align}
  \sigma (\beta) \simeq \ln \frac{1}{c_1 (\beta)} > 0 \ .
\end{align}
In the strong coupling region, this result reproduces the area law falloff in the $SU(N)$ lattice gauge model obtained in \cite{KSK20}. Moreover, by taking the continuous group limit $N \to \infty$ ,we find that the area law for $N \geqslant 4$ persists in the $U(1)$ lattice gauge model.

\begin{figure}[!h]
\centering
\begin{subfigure}{60mm}
  \centering\includegraphics[width=60mm]{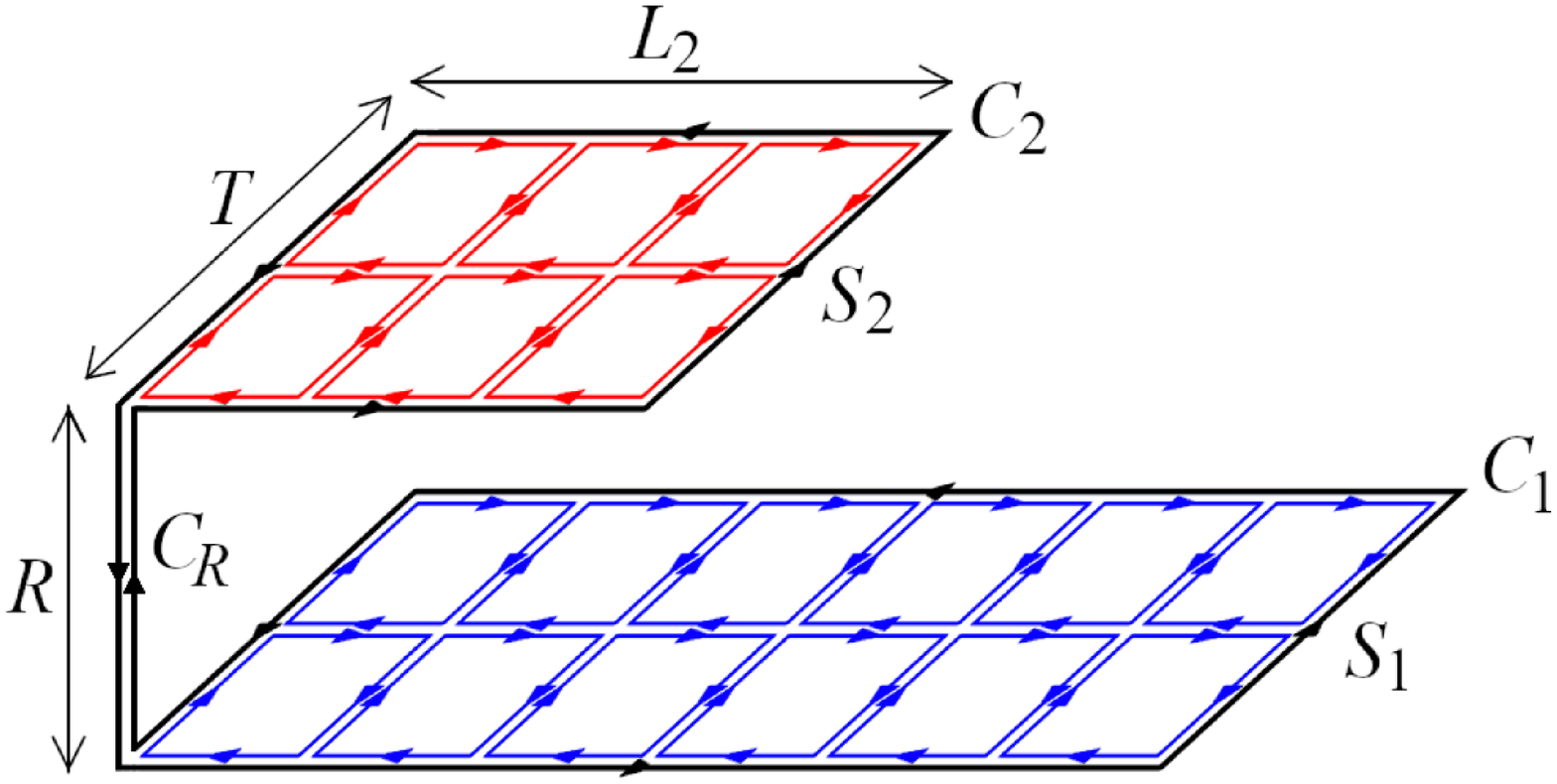}
  \hspace{0mm} \caption{$R$-independent contribution}
  \label{fig7}
\end{subfigure}
\begin{subfigure}{60mm}
  \centering\includegraphics[width=60mm]{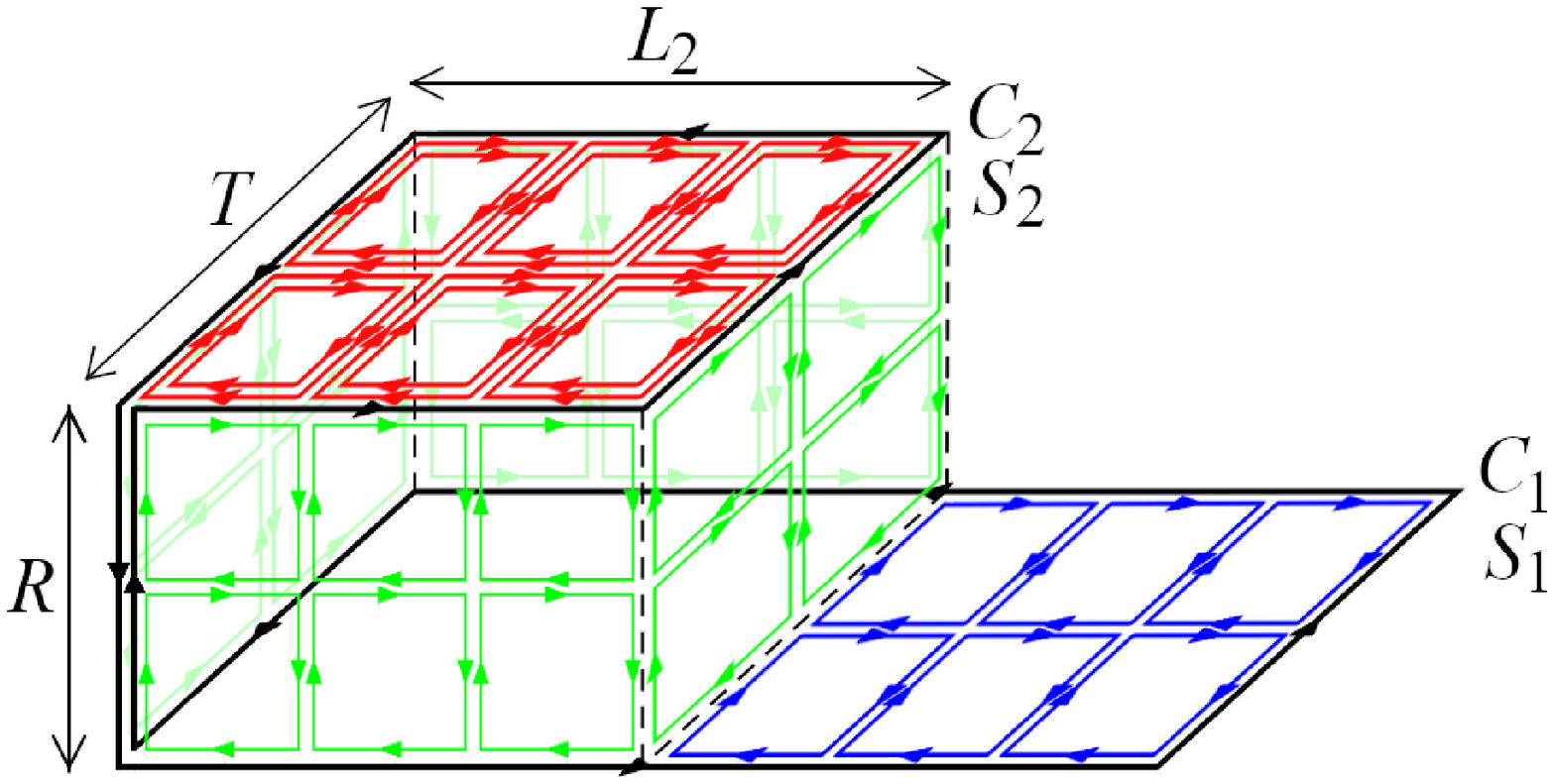}
  \hspace{0mm} \caption{$R$-dependent contribution}
  \label{fig8}
\end{subfigure}
\hspace{0mm} \caption{A shifted double-winding Wilson loop, (a) $R$-independent contribution, (b) $R$-dependent contribution}
\label{fig78}
\end{figure}

Furthermore, we also evaluate the expectation value of a shifted double-winding Wilson loop in the lattice $Z_N$ pure gauge model. The leading contribution to a shifted double-winding Wilson loop average can be given by the 2 types of tiling by a set of plaquettes, as shown in the Fig.\ref{fig78}.

%%%%%%%%%%%%%%%%%%%%%%%%%%%%%%%%%%%%%%%%%%%%%%%%%%%%%%%%%%%%

%%%%%%%%%%%%%%%%%%%%%%%%%%%%%%%%%%%%%%%%%%%%%%%%%%%%%%%%%%%%

The result of the shifted double-winding Wilson loop average up to the leading contribution is given by
\begin{align}
  {\langle W(C_1 \cup C_2) \rangle}_{R \neq 0} \simeq
  \begin{cases}
    {c_1 (\beta)}^{|S_1|+|S_2|} + {c_1 (\beta)}^{2R(L_2 + T)} \cdot {c_1 (\beta)}^{|S_1|-|S_2|} & (N=2) \\
    {c_1 (\beta)}^{|S_1|+|S_2|} + {c_1 (\beta)}^{2R(L_2 + T)} \cdot {c_1 (\beta)}^{|S_1|} & (N=3) \\
    {c_1 (\beta)}^{|S_1|+|S_2|} + {c_1 (\beta)}^{2R(L_2 + T)} \cdot {c_2 (\beta)}^{|S_2|} {c_1 (\beta)}^{|S_1|-|S_2|} & (N \geqslant 4)
  \end{cases} \ .
\end{align}
This result reproduces the $R$-dependent behavior of the shifted double-winding Wilson loop average in \cite{KSK20}. In particular, we obtain the (non-zero) mass gap from the case of $S_1 = S_2 = 1$ and $R \gg 1$ in the above result:
\begin{align}
  \Delta (\beta) = 4 \ln \frac{1}{c_1 (\beta)} > 0 \ .
\end{align}

\section{Lattice $Z_N$ gauge-scalar theory}

Next, we consider the lattice $Z_N$ gauge-scalar model with the frozen scalar field norm $R$ for simplicity. The action of this model with the coupling constants defined by $\beta := 1/ g^2$ and $K := R^2$ on a $D$-dimensional lattice $\Lambda$ with unit lattice spacing is given by
\begin{equation}
  S[U,\varphi] = \beta \sum_{p \in \Lambda} \mathrm{Re} \ U_p + K \sum_{\ell \in \Lambda} \mathrm{Re} \left( {\varphi}_{x} {U}_{\ell} {\varphi}_{x+\ell}^{*} \right ) \ ,
\end{equation}
where $\ell$ labels a link, and $p$ labels an elementary plaquette. ${U}_{\ell}$ is a $Z_N$ link variable on link $\ell$ and ${\varphi}_{x}$ is a $Z_N$ scalar field at site $x$ which transforms according to the fundamental representation of the gauge group $Z_N$.

In this model, the expectation value of an operator $\mathscr{F}$ has the form
\begin{align}
  {\langle \mathscr{F} \rangle}_{\Lambda}
    &:= {Z}_{\Lambda}^{-1} \int \prod_{\ell \in \Lambda} d{U}_{\ell} \prod_{x \in \Lambda} d{\varphi}_{x} \ {e}^{S[U,\varphi]} \mathscr{F}
      = {Z}_{\Lambda}^{-1} \int \prod_{\ell \in \Lambda} d{U}_{\ell} \ h[U] \ {e}^{\beta \sum_{p \in \Lambda} \mathrm{Re} \ U_p} \mathscr{F} \ , \notag \\ 
  {Z}_{\Lambda}
    &:= \int \prod_{\ell \in \Lambda} d{U}_{\ell} \prod_{x \in \Lambda} d{\varphi}_{x} \ {e}^{S[U,\varphi]} \ , \quad h[U] := \int \prod_{x \in \Lambda} d{\varphi}_{x} \ {e}^{ K \sum_{\ell \in \Lambda} \mathrm{Re} ( {\varphi}_{x} {U}_{\ell} {\varphi}_{x+\ell}^{*} ) } \ .
\end{align}

According to \cite{OS78}, we can perform the \textit{cluster expansion} by introducing the new variable ${\rho}_{p}$ and the new measure $d{\mu}_{\Lambda}$ which absorbs the scalar part $h[U]$:
\begin{align}
  {\langle \mathscr{F} \rangle}_{\Lambda}
    &= \frac{ \int d{\mu}_{\Lambda} \prod_{p \in \Lambda} \left( 1+ {\rho}_{p} \right) \mathscr{F} }{ \int d{\mu}_{\Lambda} \prod_{p \in \Lambda} \left( 1+ {\rho}_{p} \right) }
    = \sum_{Q(Q_0) \subset \Lambda} \int d{\mu}_{\Lambda} \ \mathscr{F} \prod_{p \in Q(Q_0)} {\rho}_{p} \cdot \frac{ {Z}_{{[Q(Q_0) \cup Q_0]}^{c}} }{ {Z}_{\Lambda} } \ , \\
  d{\mu}_{\Lambda}
    &:= \frac{ \prod_{\ell \in \Lambda} d{U}_{\ell} \ h[U] }{ \int \prod_{\ell \in \Lambda} d{U}_{\ell} \ h[U] } \ , \quad
  {\rho}_{p}
    := {e}^{\beta \mathrm{Re} \ U_p } -1 \ ,
\end{align}
where $Q_0$ is the set of plaquettes which is the support of $\mathscr{F}$ and $Q(Q_0)$ is the set of plaquettes which is connected to $Q_0$. For the general set of plaquettes $Q$, $Q^c$ represents the complement of $Q$. Here, $Z_Q$ is defined by
\begin{equation}
  {Z}_{Q} := \sum_{Q' \subset Q} \int d{\mu}_{\Lambda} \prod_{p \in Q'} {\rho}_{p} \ .
\end{equation}

\begin{figure}[!h]
  \centering\includegraphics[width=40mm]{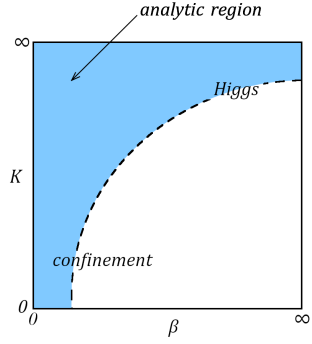}
  \caption{The analytic region on the $\beta$-$K$ plane}
  \label{fig9}
\end{figure}

Note that ${\rho}_{p} \sim \mathcal{O} (\beta)$ for $\beta \ll 1$. It has been showed in \cite{FS79} that the confinement region ($0 \leqslant \beta \ll 1 , K \ll1$) and the Higgs region ($\beta \gg 1 , K_c \leqslant K < \infty$) are analytically continued in a single “analytic region”, where the cluster expansion converges uniformly. See Fig.\ref{fig9}.

%%%%%%%%%%%%%%%%%%%%%%%%%%%%%%%%%%%%%%%%%%%%%%%%%%%%%%%%%%%%

%%%%%%%%%%%%%%%%%%%%%%%%%%%%%%%%%%%%%%%%%%%%%%%%%%%%%%%%%%%%

To evaluate $h[U]$, we apply the character expansion and perform the group integration. Ignoring the contributions from multiple plaquettes, then we obtain the expression which is valid up to the lowest plaquettes order:
\begin{align}
  h[U]
    &= \int \prod_{x \in \Lambda} d{\varphi}_{x} \prod_{\ell \in \Lambda} \left[ b_0 (K) + b_1 (K) {\varphi}_{x} {U}_{\ell} {\varphi}_{x+\ell}^{*} + \cdots + {b}_{N-1} (K) { \left( {\varphi}_{x} {U}_{\ell} {\varphi}_{x+\ell}^{*} \right) }^{N-1} \right] \notag \\
    &= {N}^{|\Lambda|} {b_0 (K)}^{D|\Lambda|} \prod_{p \in \Lambda} \sum_{n=0}^{N-1} {c_n (K)}^{4} U_p^n + \cdots \ .
\end{align}

We estimate the leading contribution to the double-winding Wilson loop average with the above $h[U]$, we also apply the character expansion for ${\rho}_{p}$ and evaluate the upper bound of the cluster expansion by using the binominal expansion. We find that there is an correspondence between the evaluation for the $Z_N$ lattice gauge model and for the estimated upper bound for the $Z_N$ lattice gauge-scalar model:
\begin{align}
  c_n (\beta) \mapsto a_n (\beta , K) := &\frac{ \left[ b_0 (\beta) - {e}^{\beta} \right] {c_n (K)}^{4} + b_1 (\beta) {{c}_{n+1} (K)}^{4} + \cdots + {b}_{N-1} (\beta) {{c}_{N+n-1} (K)}^{4} }{ b_0 (\beta) + b_1 (\beta) {c_1 (K)}^{4} + \cdots + {b}_{N-1} (\beta) {{c}_{N-1} (K)}^{4} } + {c_n (K)}^{4} \ . \notag \\
  &(\hspace{-2mm} \mod N \ , \ n = 1, \cdots , N-1) 
\end{align}
Note that $a_n (\beta, 0) = c_n (\beta)$ and $a_n (\beta, \infty) = 1$. The above estimation is valid only for the values of parameter $\beta$ and $K$ on the analytic region in the range where the string breaking does not occur. 

By applying the same method as the above, we obtain the estimation for the coplanar double-winding Wilson loop average:
\begin{align}
  {\langle W(C_1 \cup C_2) \rangle}_{R=0} \lesssim
  \begin{cases}
    {a_1 (\beta,K)}^{|S_1|-|S_2|} & (N=2) \\
    {a_1 (\beta,K)}^{|S_1|} & (N=3) \\
    {a_2 (\beta,K)}^{|S_2|} {a_1 (\beta,K)}^{|S_1|-|S_2|} & (N \geqslant 4)
  \end{cases}
\end{align}
and we obtain the (non-zero) string tension from the above result:
\begin{align}
  \sigma (\beta,K) \gtrsim \ln \frac{1}{a_1 (\beta,K)} > 0 \ .
\end{align}

This estimation suggests that the area law falloff in the $Z_N$ lattice gauge model persists in the $Z_N$ lattice gauge-scalar model and the $K \to 0$ limit agrees with the pure gauge case. Moreover, for $\sigma (\beta , K)$, the $K \to 0$ limit agrees with $\sigma (\beta)$ in the $Z_N$ lattice gauge model, and $K \to \infty$ limit converges to 0 uniformly in $\beta$. In other words, the string tension is non-zero on the analytic region.

Additionally, we also estimate the shifted double-winding Wilson loop average:
\begin{align}
  &{\langle W(C_1 \cup C_2) \rangle}_{R \neq 0} \notag \\
  &\quad \lesssim
  \begin{cases}
    {a_1 (\beta,K)}^{|S_1|+|S_2|} + {a_1 (\beta,K)}^{2R(L_2 + T)} \cdot {a_1 (\beta,K)}^{|S_1|-|S_2|} & (N=2) \\
    {a_1 (\beta,K)}^{|S_1|+|S_2|} + {a_1 (\beta,K)}^{2R(L_2 + T)} \cdot {a_1 (\beta,K)}^{|S_1|} & (N=3) \\
    {a_1 (\beta,K)}^{|S_1|+|S_2|} + {a_1 (\beta,K)}^{2R(L_2 + T)} \cdot {a_2 (\beta,K)}^{|S_2|} {a_1 (\beta,K)}^{|S_1|-|S_2|} & (N \geqslant 4)
  \end{cases}
\end{align}
and we obtain the (non-zero) mass gap from the case of $S_1 = S_2 = 1$ and $R \gg 1$ in the above result:
\begin{align}
  \Delta (\beta,K) \gtrsim 4 \ln \frac{1}{a_1 (\beta,K)} > 0 \ .
\end{align}

For $\Delta (\beta , K)$, the $K \to 0$ limit agrees with $\Delta (\beta)$ in the $Z_N$ lattice gauge model, and $K \to \infty$ limit converges to 0 uniformly in $\beta$. In other words, the mass gap is non-zero on the analytic region. 

\section{Conclusion}

We investigated the area law falloff of the double-winding Wilson loops in the $Z_N$ lattice gauge model and $Z_N$ lattice gauge-scalar model, where the gauge group is the center group of the original $SU(N)$.
First, we evaluated the $N$-dependent area law falloff for the coplanar double-winding Wilson loop average up to the leading contribution. We found the $N$-dependence of the area law falloff in the $Z_N$ lattice gauge model, which reproduces the area law
falloff in the $SU(N)$ lattice gauge model obtained in \cite{KSK20}.
Secondly, we also checked the limit $N \to \infty$, the area law falloff for $N \geqslant 4$ persists in the $U(1)$ lattice gauge model. This result implies that the coplanar double-winding Wilson loop average in the $U(N)$ lattice gauge model and the $SU(N) (N \geqslant 4)$ lattice gauge model obeys the same area law up to the leading contribution.
Furthermore, we also considered the shifted double-winding Wilson loop average up to the leading contributions. This result reproduces the $R$-dependent behavior in the $SU(N)$ lattice gauge model obtained in \cite{KSK20}. We obtained the (non-zero) mass gap $\Delta (\beta)$ from this result.
Finally, we extended the above study for the $Z_N$ lattice gauge-scalar model on the analytic region. We found that the
area law falloff in the $Z_N$ lattice gauge model persists in the $Z_N$ lattice gauge-scalar model. We discovered that the string tension and the mass gap are non-zero on the analytic region from this estimation.

\end{document}